\documentclass[fleqn,twoside]{article}
\usepackage{amssymb,amsfonts,espcrc2}
\usepackage{graphicx}

\newcommand{\pslush}{p\hspace{-0.16cm}/}
\newcommand{\ind}[2]{^{#1}_{\mbox{\scriptsize #2}}}
\newcommand{\al}[2]{\alpha\ind{#1}{#2}}
\newcommand{\ro}[1]{\rho^{(#1)}}

\def\KL{K\"all\'en--Lehmann }

\def\sfa{\mbox{\sf A}}
\def\bz{\beta_0}
\def\fpb{\frac{4 \pi}{\bz}}
\def\mpi{m_{\pi}}
\def\epem{e^{+}e^{-}}
\def\alc{\alpha_{\mbox{\tiny C}}}
\def\nf{n_{\mbox{\scriptsize f}}}

\title{The QCD analytic running coupling and chiral symmetry breaking}

\author{A.V.~Nesterenko\address[UV]{Departamento de F\'\i sica Te\'orica
and IFIC, Centro Mixto, Universidad de Valencia-CSIC,
E-46100, Burjassot, Valencia, Spain}\address[JINR]{Bogoliubov Laboratory 
of Theoretical Physics, JINR, Dubna, 141980, Russian Federation}
and
J.~Papavassiliou\addressmark[UV]}

\begin{document}

\begin{abstract}  
We study the dependence on the pion mass of the QCD effective charge by
employing the dispersion relations for the Adler $D$~function. This new
massive analytic running coupling is compared to the effective coupling
saturated by the dynamically generated gluon mass. A qualitative picture
of the possible impact of the former coupling on the chiral symmetry
breaking is presented. \vskip-2.5mm 
\end{abstract}

\maketitle

     The basic idea behind the analytic approach to Quantum Field
Theory is to supplement the perturbative treatment of the
renormalization group (RG) formalism with the nonperturbative
information encoded in the corresponding 
dispersion relations~\cite{AQED}. The
latter, being based on the ``first principles'' of the theory, 
provide one with the
definite analytic properties in the kinematic variable of a physical
quantity at hand~\cite{ShSol,PRD1}. In practice the analytization
procedure \cite{ShSol} amounts to the restoration of the correct
analytic properties for a given quantity $\sfa(q^2)$ by imposing the
\KL representation
\begin{equation}
\label{DefAn}
\Bigl\{\sfa(q^2)\Bigr\}_{\mbox{$\!$\scriptsize an}} =
\int_{0}^{\infty} \!\frac{\varrho(\sigma)}{\sigma+q^2}\,
d \sigma.
\end{equation}
Here the spectral function $\varrho(\sigma)$ can be defined by
the initial (perturbative) expression for~$\sfa(q^2)$:
\begin{equation}
\varrho(\sigma) = \frac{1}{2\pi i}\, \lim_{\varepsilon \to 0_{+}}
\Bigl[\sfa(-\sigma-i\varepsilon)-\sfa(-\sigma+i\varepsilon)\Bigr].
\end{equation}

     A distinctive feature of the model for the QCD analytic
invariant charge $\al{}{an}(q^2)$~\cite{PRD1} is
the application of the analytization procedure~(\ref{DefAn}) to the
perturbative expansion of the RG $\beta$ function:
\begin{equation}
\label{AnRGEq}
\frac{d\,\ln a\ind{(\ell)}{an}(\mu^2)}{d\,\ln \mu^2} =
- \left\{\sum_{j=0}^{\ell-1} \frac{\beta_{j}}{\bz^{j+1}}
\Bigl[a\ind{(\ell)}{s}(\mu^2)
\Bigr]^{j+1}\!\right\}_{\!\!\mbox{\scriptsize an}}\!\!.
\end{equation}
Here $\al{(\ell)}{s}(q^2)$ denotes the $\ell$-loop perturbative
running coupling, $a(q^2)=\alpha(q^2)\bz/(4\pi)$, and 
$\beta_j$~is the $\beta$ function expansion coefficient. At the
one-loop level the renormalization group equation~(\ref{AnRGEq}) 
can be solved explicitly:
\begin{equation}
\label{AIC1L}
\al{(1)}{an}(q^2) = \fpb \, \frac{z - 1}{z \, \ln z},
\qquad z = \frac{q^2}{\Lambda^2},
\end{equation}
where $q^2>0$ stands for the spacelike momentum. The solution to
Eq.~(\ref{AnRGEq}) can also be represented in the form of
the \KL integral
\begin{equation}
\label{AICHLKL}
\al{(\ell)}{an}(q^2) = \fpb \int_{0}^{\infty}
\frac{\ro{\ell}(\sigma)}{\sigma + z}\, d \sigma,
\end{equation}
where the one-loop spectral density is
\begin{equation}
\label{SpDns1L}
\ro{1}(\sigma) =
\left(1+\frac{1}{\sigma}\right)\frac{1}{\ln^2\!\sigma+\pi^2},
\end{equation}
and the explicit expression for the $\ell$-loop $\ro{\ell}(\sigma)$
can be found in Ref.~\cite{PRD1}.

\begin{figure}[t]
\includegraphics[width=60mm]{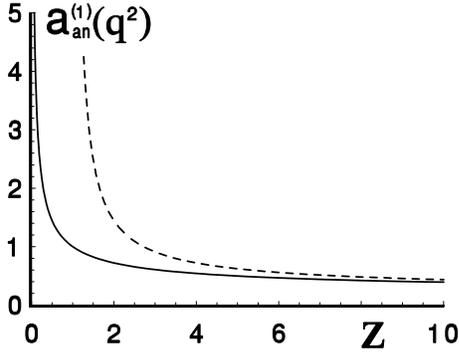}
\vskip-7mm
\caption{The one-loop perturbative coupling (dashed curve)
and the analytic charge (Eq.~(\protect\ref{AIC1L}))
(solid curve), $a(q^2)=\alpha(q^2)\bz/(4\pi), \;
z=q^2/\Lambda^2$.}
\label{fig:1}
\vskip-5mm
\end{figure}

     The model (\ref{AIC1L}) shares all the advantages of the
analytic approach: it contains no unphysical singularities at $q^2>0$
and possesses good higher loop and scheme stability. Besides, it has
proved to be successful in description of hadron dynamics of the both
perturbative and intrinsically nonperturbative nature. It is worth
noting that the massless analytic effective charge~(\ref{AICHLKL})
incorporates the ultraviolet asymptotic freedom with the infrared
enhancement (i.e., the singular behavior at $q^2=0$) in a single
expression, see Figure~\ref{fig:1}.

     In this talk we will outline how the behavior of the running
coupling (\ref{AICHLKL}) is affected by the pion mass entering the
Adler $D$ function, thus giving rise to an infrared finite value for 
this coupling. We will also argue that this analytic effective charge 
may be relevant to the study of chiral symmetry breaking (CSB) through 
the Schwinger-Dyson equations.

     A certain insight into the nonperturbative aspects of the strong
interaction can be provided by the Adler $D$~function~\cite{Adler}
\begin{equation}
\label{AdlerDef}
D(q^2) = \frac{d\, \Pi(q^2)}{d \ln q^2},
\end{equation}
with $\Pi(q^2)$ being hadronic vacuum polarization function. In
particular, Eq.~(\ref{AdlerDef}) is related to the measurable
ratio $R(s)$ of the $\epem$ annihilation into hadrons through
the dispersion relation~\cite{Adler}
\begin{equation}
\label{AdlerDisp}
D(q^2) = q^2 \int_{4\mpi^2}^{\infty}
\frac{R(s)}{(s + q^2)^2} \, d s.
\end{equation}
In turn, this equation implies the definite analytic properties in
the $q^2$ variable for $D(q^2)$: it is an analytic function in the
complex $q^2$-plane with the only cut beginning at the two--pion
threshold $-\infty < q^2 \le -4\mpi^2$ along the negative semiaxis of
real~$q^2$.

     In the ultraviolet domain the Adler $D$ function is usually
computed in the framework of perturbation theory:
\begin{equation}
\label{Adler}
D(q^2) = 3 \sum_{f} Q_f^2 \left[1 + d(q^2) \right],
\end{equation}
where $Q_f$ is the charge of the $f$-th quark,
\begin{eqnarray}
\label{AdlerPert}
d(q^2) \simeq
d_1\left[\frac{\al{}{s}(q^2)}{\pi}\right]   +
d_2\left[\frac{\al{}{s}(q^2)}{\pi}\right]^2 + \ldots,
\end{eqnarray}
$d_1 = 1$, $d_2 \simeq 1.9857 - 0.1153\,\nf$, and $\nf$ is the
number of active quarks.
However, such approximation of Eq. (\ref{AdlerPert}) violates
the analyticity condition that $D(q^2)$ must satisfy, due to the
spurious singularities of the perturbative running 
coupling~$\al{}{s}(q^2)$. Nevertheless, this difficulty,
which is an artifact of the perturbative treatment,
can be eliminated by imposing the analyticity requirement of the form
\begin{equation}
\label{KLM}
d(q^2, \mpi^2) = \int_{4 \mpi^2}^{\infty}
\frac{\varkappa(\sigma)}{\sigma + q^2} \, d \sigma
\end{equation}
on the right hand-side of Eq.~(\ref{AdlerPert}). Therefore, the QCD
effective charge itself has to satisfy the integral representation
\begin{equation}
\label{AICM}
\al{(\ell)}{an}(q^2, \mpi^2) = \fpb \!\int_{\chi}^{\infty}\!
\frac{\ro{\ell}(\sigma)}{\sigma + z} \, d \sigma, \;\;
\chi=\frac{4\mpi^2}{\Lambda^2},
\end{equation}
where $\ro{\ell}(\sigma)$ is the $\ell$-loop spectral 
density~\cite{PRD1,Paris04}.

\begin{figure}[t]
\includegraphics[width=60mm]{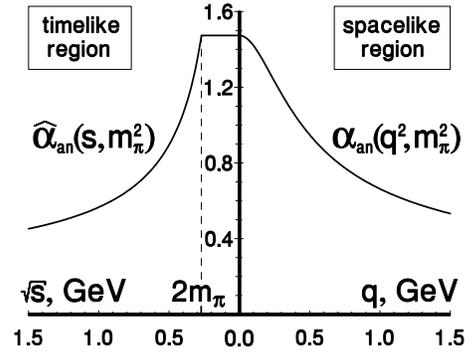}
\vskip-7mm
\caption{The one--loop massive analytic effective charge 
(\protect\ref{AICM}) in the spacelike and timelike domains. 
The values of parameters are: $\nf=2$ active quarks, 
$\Lambda = 623\,$MeV.}
\label{fig:2}
\vskip-5mm
\end{figure}

     The behavior of the one-loop massive analytic charge~(\ref{AICM}) in
the spacelike and timelike infrared domains is shown in Figure~\ref{fig:2}.
The value of $\Lambda=(623 \pm 81)\,$MeV is found by making use 
of the experimental data on the inclusive $\tau$~lepton decay~\cite{AICM}. 
It is worth
noting that the nonvanishing pion mass drastically affects the
infrared behavior of the analytic coupling in hand: instead of the
enhancement in the massless case~(\ref{AICHLKL}) one has here a finite
infrared limiting value of the effective charge~(\ref{AICM}) (see also
Refs.~\cite{Paris04,AICM} for the details).

     Based on the study of the {\it gauge invariant} Schwinger-Dyson
equations, Cornwall proposed a long time ago that the self-interactions
of gluons give rise to a dynamical gluon mass, while preserving at
the same time the local gauge symmetry of the theory
\cite{Cornwall:1982zr}. This gluon ``mass'' is not a directly
measurable quantity, but must be related to other physical
parameters, such as the glueball spectrum, the energy needed to pop
two gluons out of the vacuum, the QCD string tension, or the QCD
vacuum energy.

     One of the main phenomenological implications of this
analysis is that the presence of the gluon mass $m_g$ saturates the 
running of the strong coupling, forcing it to ``freeze'' in the 
infrared domain. In particular, the nonperturbative effective coupling 
obtained in Ref.~\cite{Cornwall:1982zr} is given by
\begin{equation}
\label{JMC}
\alc(q^2)= \fpb
\frac{1}{\ln\!\left[z + 4M_g^2(q^2)/\Lambda^2 \right]},
\end{equation}
where $M_g(q^2)$ is the dynamical gluon mass
\begin{equation}
\label{runmass}
M^2_g(q^2) = m_g^2 \left[\frac{\ln\left(z + 4 m_g^2/\Lambda^2\right)}
{\ln\left(4 m_g^2/\Lambda^2\right)}\right]^{-12/11}.
\end{equation}
The coupling (\ref{JMC}) has the infrared finite limiting value
$\alc(0)= 4\pi \left[ \bz \ln (4 m_g^2/\Lambda^2)\right]^{-1}$. For a
typical values of $m_g = 500\,$MeV and $\Lambda = 300\,$MeV, one
obtains for the case of pure gluodynamics ($\nf=0$) an estimation
$\alc(0) \simeq 0.5$. An independent analysis \cite{Cornwall:1989gv} 
yields a maximum
allowed value for $\alc(0)$ of about~0.6. The incorporation of
fermions into the effective charge \cite{Papavassiliou:1991hx} 
\begin{equation}
\label{JPRC}
\al{}{cp}(q^2)=\frac{4\pi}{11\ln(z+\chi_g)-2\nf\ln(z+\chi_q)/3}
\end{equation}
does not change the picture qualitatively (at least for quark masses 
of the order of~$\Lambda$). In equation~(\ref{JPRC}) 
$\chi_g=4m_g^2/\Lambda^2$, $\chi_q=4m_q^2/\Lambda^2$,
$m_g=(500 \pm 100)\,$MeV stands for the gluon mass and $m_q=350\,$MeV
is a light quark constituent mass.

     The effective coupling of Eq.~(\ref{JMC}) was the focal
point of extensive scrutiny, and has been demonstrated to furnish a
unified description of a wide variety of the low energy QCD data
\cite{Aguilar:2004td}.

     However, an important unresolved question in this context is the
incorporation of the QCD effective charge into the standard
Schwinger-Dyson equation governing the dynamics of the quark
propagator $S(p)$
\begin{equation}
\label{gap}
S^{-1}(p) = S_{0}^{-1}(p) - g^2 \int \frac{d^4k}{(2\pi)^4}
\gamma_{\mu} \, S \,\Gamma_{\nu} \,\Delta^{\mu\nu}.
\end{equation}

\begin{figure}[t]
\includegraphics[width=60mm]{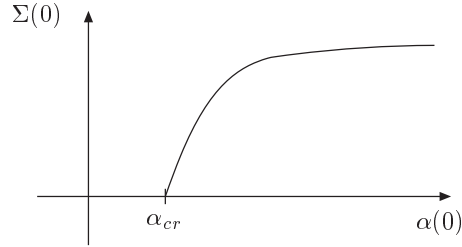}
\vskip-7mm
\caption{A typical dependence of $\Sigma(0)$ on the limiting value 
$\alpha(0)$, $\,S^{-1}(p) = A(p)\pslush + \Sigma(p)$.}
\label{fig:3}
\vskip-5mm
\end{figure}

     Specifically, since QCD is not a fixed point theory, the usual
QED-inspired gap equation must be modified, in order to incorporate the
running charge and asymptotic freedom. The usual way of accomplishing
this eventually boils down to the replacement $1/k^2 \to
\alpha(k^2)/k^2$ in the corresponding kernel of the gap equation, where
$\alpha(k^2)$ is the QCD running coupling. The inclusion of $\alpha(k^2)$
is essential for arriving at an integral equation for $S(p)$ which is
well-behaved in the ultraviolet. However, since
the perturbative form of $\alpha(k^2)$ diverges at low energies as
$1/\ln(k^2/\Lambda^2)$ when $k^2 \to \Lambda^2$, some form of the
infrared regularization for $\alpha(k^2)$ is needed, whose details depend
on the specific assumptions one is making regarding the
nonperturbative hadron dynamics. At this point the issue of the
critical coupling makes its appearance. Specifically, as is
well-known, there is a critical infrared limiting value of the
running coupling, to be denoted by $\al{}{cr}$, below which there are
no nontrivial solutions to the resulting gap equation, i.e., there is
no CSB, see Fig.~\ref{fig:3}.

     The incorporation of the effective charge of Eq.~(\ref{JMC})
into a gap equation has been studied for the first time in
Ref.~\cite{Haeri:1990yj}. There it was concluded that CSB solutions
for $\Sigma(p)$ could be obtained only for unnaturally small values
of the gluon mass, namely $m_g/\Lambda \simeq 0.8$. This is so
because the typical value of $\al{}{cr}$ found in the standard
treatment of the gap equation is $\al{}{cr} \simeq 1.2$, which is
what the expression for $\alc(0)$ yields for the above value
of~$m_g/\Lambda$. This issue was further investigated in
Ref.~\cite{Papavassiliou:1991hx}, where a system of coupled gap and
vertex equations was considered. The upshot of this study was that no
consistent solutions to the system of integral equations could be
found, due to the fact that the allowed values for~$\alpha(0)$,
dictated by the vertex equation, were significantly lower than
$\alc(0)$, i.e., not large enough to trigger chiral symmetry breaking.

\begin{figure}[t]
\includegraphics[width=60mm]{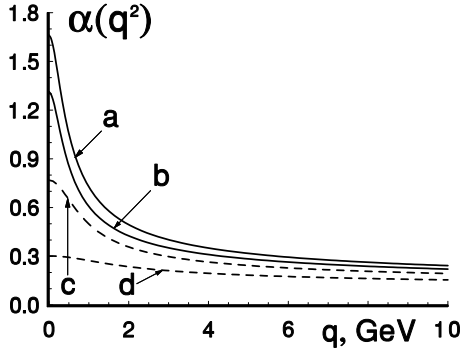}
\vskip-7mm
\caption{Comparison of the massive analytic running coupling
(Eq.~(\protect\ref{AICM}), solid curves) with the effective 
charge~(\protect\ref{JPRC})
(dashed curves). The values of parameters are: $\nf=2$ active quarks,
$\Lambda = 704\,$MeV (a), $\Lambda = 542\,$MeV (b),
gluon mass $m_g=400\,$MeV and $\Lambda = 350\,$MeV (c),
gluon mass $m_g=600\,$MeV and $\Lambda = 150\,$MeV~(d).}
\label{fig:4}
\vskip-7mm
\end{figure}

     In what follows we will suggest a possible resolution of this
problem. The basic observation is captured in Figure~\ref{fig:4}:
the effective charge with a gluon mass (dashed curves) and the analytic
charge (\ref{AICM}) (solid curves) coincide for a large range of
momenta, and they only begin to differ appreciably in the deep
infrared domain ($k^2 \lesssim \Lambda^2$). In this region the analytic 
charge (\ref{AICM}) rises abruptly, almost doubling its size between 
$k^2=\Lambda^2$ and $k^2=0$, whereas the running coupling~(\ref{JPRC}) 
in the same momentum interval
remains essentially fixed to a value of about~0.6. A possible picture
that seems to emerge from this observation is the following. It may
be that the concept of the dynamically
generated gluon mass fails to capture all the relevant dynamics in
the very deep infrared, where confinement or other nonperturbative
effects make their appearance. At that point it may be preferable to
switch to a description in terms of the analytic charge (\ref{AICM}),
which (i)~coincides with that of Cornwall in the region where the
latter furnishes a successful description of data, and (ii)~in
addition, because it overcomes the critical value $\al{}{cr}$,
offers the possibility of accounting for CSB at the level of gap
equations. It would be interesting to carry out a detailed study of
the gap equation, with the analytic charged plugged into, in order to
verify if indeed one encounters nontrivial solutions, whose size is
phenomenologically relevant, and if a reasonable value of the
pion-decay constant $f_{\pi}$ may be obtained.

\section*{Acknowledgments}
     Authors thank Professors D.V.\ Shirkov, A.E.\ Dorokhov, I.L.\ Solovtsov,
and N.~Stefanis for the stimulating discussions. J.P.~thanks the organizers 
of QCD~04 for their hospitality. The work has been supported by grants 
RFBR (02-01-00601, 04-02-81025), NS-2339.2003.2, and CICYT FPA20002-00612.


\begin{thebibliography}{11}

\bibitem{AQED} P.J.~Redmond, Phys.\ Rev.\ {\bf 112}, 1404 (1958);
         N.N.~Bogoliubov, A.A.~Logunov, and D.V. Shirkov,
         Sov.\ Phys.\ JETP {\bf 37}, 574 (1960).

\bibitem{ShSol} D.V.~Shirkov and I.L.~Solovtsov,
         Phys.\ Rev.\ Lett.\ {\bf 79}, 1209 (1997).

\bibitem{PRD1} A.V.~Nesterenko, Phys.\ Rev.\ D {\bf 62}, 094028 (2000);
         Phys.\ Rev.\ D {\bf 64}, 116009 (2001);
         Int.\ J.\ Mod.\ Phys.\ A {\bf 18}, 5475 (2003).

\bibitem{Adler} S.L.~Adler, Phys.\ Rev.\ D {\bf 10}, 3714 (1974).

\bibitem{Paris04} A.V.~Nesterenko and J.~Papavassiliou, \\
         arXiv:hep-ph/0409220.

\bibitem{AICM} A.V.~Nesterenko and J.~Papavassiliou, in preparation.

\bibitem{Cornwall:1982zr}
J.M.~Cornwall,
Phys.\ Rev.\ D {\bf 26}, 1453 (1982).

\bibitem{Cornwall:1989gv}
J.M.~Cornwall and J.~Papavassiliou,
Phys.\ Rev.\ D {\bf 40}, 3474 (1989).

\bibitem{Papavassiliou:1991hx}
J.~Papavassiliou and J.M.~Cornwall,
Phys.\ Rev.\ D {\bf 44}, 1285 (1991).

\bibitem{Aguilar:2004td}
A.C.~Aguilar, A.~Mihara and A.A.~Natale,
Int.\ J.\ Mod.\ Phys.\ A {\bf 19}, 249 (2004).

\bibitem{Haeri:1990yj}
B.~Haeri and M.B.~Haeri,
Phys.\ Rev.\ D {\bf 43}, 3732 (1991).

\end{thebibliography}
\end{document}